\documentclass[%
unsortedaddress,
preprint,
 amsmath,amssymb,
 aps,
]{revtex4-2}

\usepackage{graphicx}
\usepackage{dcolumn}
\usepackage{bm}

\usepackage{xcolor}
\begin{document}


\title{Influence of doping level on Brillouin oscillations in GaAs}

\author{Adam Dodson$^\dagger$}
\author{Andrey Baydin$^\dagger$}%
 \email{andrey.baydin@vanderbilt.edu}
 \altaffiliation[current address: ]{Department of Electrical and Computer Engineering, Rice University, Houston, Texas 77005, USA}
\author{Hongrui Wu}
\affiliation{Department of Physics and Astronomy, Vanderbilt University, Nashville, TN 37235, USA}
\author{Halina Krzyzanowska}
\affiliation{Department of Physics and Astronomy, Vanderbilt University, Nashville, TN 37235, USA}
\affiliation{Institute of Physics, Maria Curie-Sklodowska University, Pl. M. Cuire-Sklodowskiej 1, 20-031 Lublin, Poland}
\author{Norman Tolk}
\affiliation{Department of Physics and Astronomy, Vanderbilt University, Nashville, TN 37235, USA}%
\begin{abstract}
Time-domain Brillouin scattering has proved to be a unique tool for determining depth dependent material properties. Here, we show the influence of doping level in GaAs on Brillouin oscillations. Measurements were performed on intrinsic, n-type and p-type GaAs samples. The results show high sensitivity of the amplitude of Brillouin oscillations to the doping concentration. The theoretical calculations are in a good agreement with the experimental data. This work provides an insight into the specific dopant profiling as a function of depth.
\end{abstract}

\maketitle


\section{\label{sec:intro}Introduction}

Time-domain Brillouin scattering (TDBS) is an ultrafast pump-probe experimental technique based on generation and detection of coherent acoustic phonons (CAPs). The inhomogeous absorption of a pump pulse ($<$1ps) in a material generates a CAP or strain wave of the order of nanometers that traverses the material \cite{Thomsen1986}. The travelling CAP wave locally perturbs optical constants of the host. Thus, the reflection and transmission of an optical probe pulse that is delayed with respect to the pump pulse is modulated by the CAP wave. The interference of the probe light waves reflected from the surface of the material and the traveling CAP wave results in an oscillatory time dependent signal referred as Brillouin oscillations. Their amplitude, frequency and decay are highly sensitive to optical and elastic properties of a material that, in principle, can be depth dependent \cite{Gusev2018}.

TBDS has proven to be invaluable tool to study different depth dependent properties of materials \cite{Gusev2018,baydin2019post} such as elastic and optical inhomogeneities in disordered films \cite{Mechri2009, Gusev2011, Lomonosov2012}, ion implantation induced modification of interfacial bonding \cite{Tas1998}, sub-$\mu$m textures in materials compressed at megabar pressures \cite{Nikitin2015, Kuriakose2016}, doping profiles \cite{Hudert2008}, depth-dependent stress \cite{Dai2016}, imaging of grain microstructure \cite{Khafizov2016, WANG201934}, and determination of laser-induced temperature gradients in liquids \cite{Chaban2017}. It has been shown that TDBS is sensitive to ion implantation induced damage in gallium arsenide \cite{Steigerwald2009a, Steigerwald2012a, Baydin2017}, diamond \cite{Gregory2012} and silicon carbide \cite{Baydin_2016} at low fluences. 

In this paper, we report on the influence of doping level in GaAs on Brillouin oscillations. Our experiment was carried out at probe energies near the band gap of GaAs for three samples: intrinsic, n-type and p-type. The experimental results are also compared to the theoretical model developed in our earlier study \cite{Baydin-energy-dep-2019}. Previously published research on the influence of doping profiles on CAP detection and generation shows that the doping profile, the position of the interface between the differently doped regions and the thickness of the transition region can be determined \cite{Hudert2008}. Contrary to this report, we investigate the probe energy region near the band gap of GaAs and show qualitatively different way of determining doping concentrations. While another report has been shown that phase of Brillouin oscillations is sensitive to the doping level \cite{babilotte2007physical}, we focus mainly on the amplitude of Brillouin oscillations. Moreover, we discuss that TDBS can be sensitive both to the doping concentration and the type of doping.

\section{\label{sec:results}Results and Discussion}

The molecular beam epitaxy grown samples of GaAs (100) used in the experiment were purchased from the Institute of Electronic Materials Technology, Warsaw, Poland. Total of three samples were studied: intrinsic, n-type, and p-type. The doping densities for n-type (Te) and p-type (Zn) were $1.28-2.35\times10^{18}$ cm$^{-3}$ and $2.94\times10^{19}$ cm$^{-3}$, respectively. A Ti layer (19 nm) was deposited onto all samples using an e-beam evaporator for efficient generation of CAPs. Ti was chosen due to its acoustic impedance that matches one of GaAs with less than $10\%$ mismatch and, therefore, acoustic reflections are suppressed at Ti and GaAs interface. 

TDBS experiments were performed in a standard time-resolved pump-probe setup in a reflection geometry. A Coherent Mira 900 with 150-fs pulses at 76 MHz was used as a laser source. Wavelength of the laser was varied between 825 nm and 900 nm. Both beams were focused onto the specimen with spot diameters of 100 $\mu$m and 80 $\mu$m for pump and probe, respectively. The pump beam was chopped using a Thorlabs optical chopper operating at 3 kHz.

\begin{figure*}
\includegraphics[width=\textwidth]{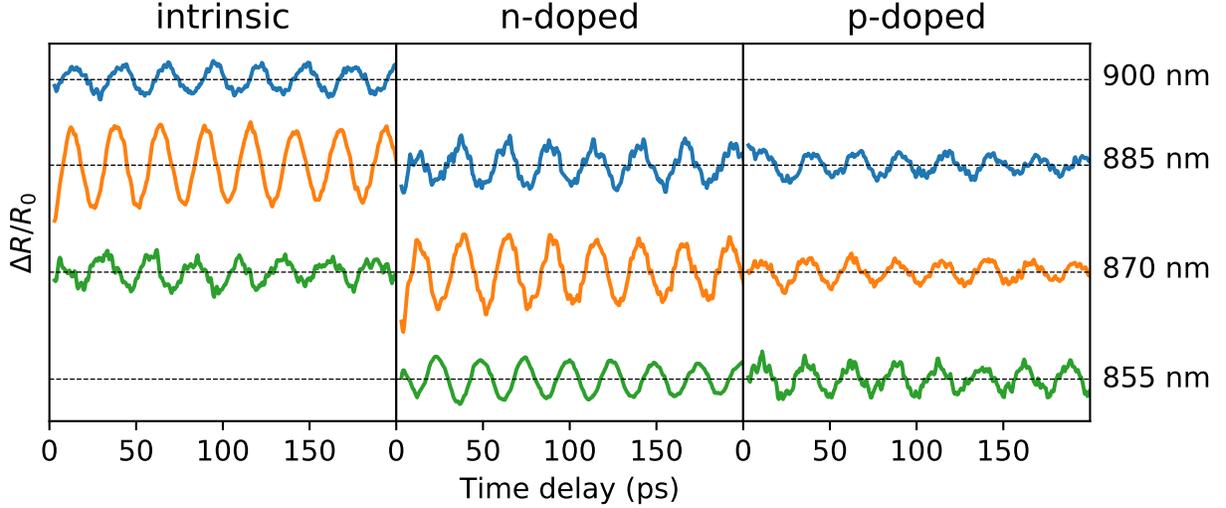}
\caption{\label{fig:raw} Brillouin oscillations for intrinsic, n-type Te-doped ($1.28-2.35\times10^{18}$ cm$^{-3}$), and p-type Zn-doped ($2.94\times10^{19}$ cm$^{-3}$) GaAs at different probe wavelengths.}
\end{figure*}

\begin{figure}
    \centering 
    \includegraphics[width=0.45\textwidth]{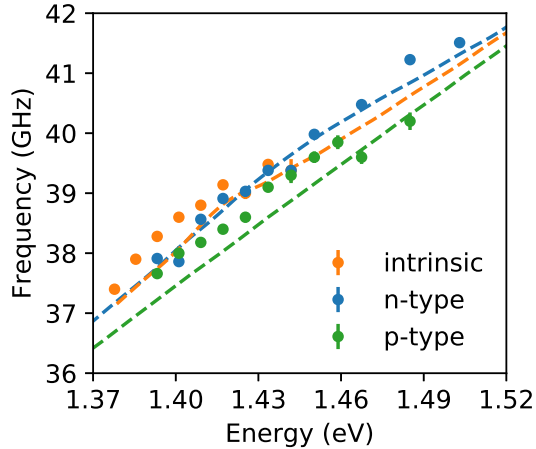}
    \caption{Frequency of Brillouin oscillations as a function of probe wavelength for intrinsic, n-type Te-doped ($1.28-2.35\times10^{18}$ cm$^{-3}$), and p-type Zn-doped ($2.94\times10^{19}$ cm$^{-3}$) GaAs. The dashed lines represent theoretical calculations. The error bars are of the order of the marker size.}
    \label{fig:freq}
\end{figure}

Figure \ref{fig:raw} shows representative Brillouin oscillations measured for intrinsic, n-type Te-doped ($1.28-2.35\times10^{18}$ cm$^{-3}$), and p-type Zn-doped ($2.94\times10^{19}$ cm$^{-3}$) GaAs at several probe wavelengths. The thermal background due to excited carriers in the Ti layer was subtracted out. The differences in experimental spectra for three samples are clearly seen in the amplitude of Brillouin oscillations. The amplitude changes as a function of probe wavelength for intrinsic and n-type GaAs, while it is constant for p-type GaAs in the probed wavelength range. The frequency of Brillouin oscillations for the studied samples is shown in Figure \ref{fig:freq}. The frequency dependence is well described by the equation $f=2\sqrt{n^2-sin^2\theta} v E_{probe}$, where $\theta$ is an angle of incidence of the probe beam (30$^\circ$), $v$ is the longitudinal speed of sound, $n$ is the index of refraction of GaAs, and $E_{probe}$ is the probe energy. The index of refraction was measured by ellipsometry for all samples. The changes in the frequency with respect to doping concentration are negligible and will not be discussed further in this paper. 

\begin{figure}
    \centering
    \includegraphics[width=0.45\textwidth]{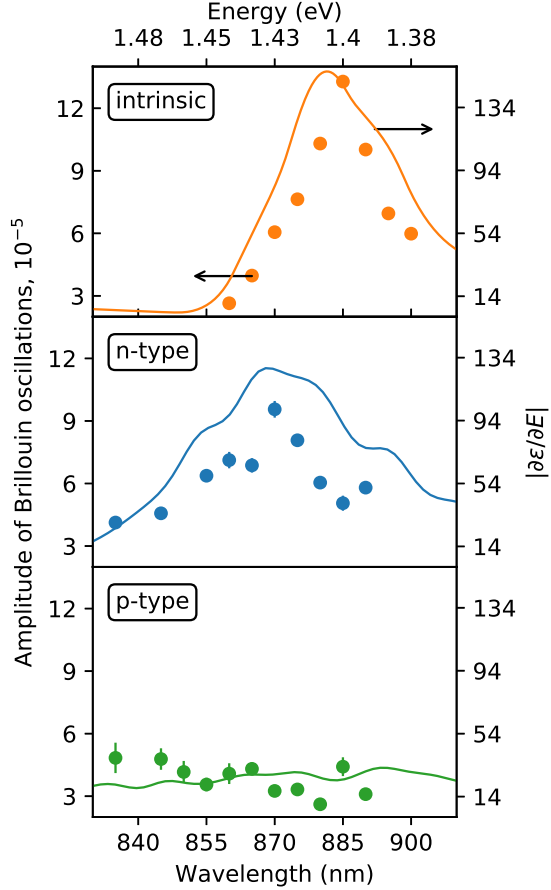}
    \caption{Amplitude of Brillouin oscillations as a function of probe wavelength for intrinsic, n-type Te-doped ($1.28-2.35\times10^{18}$ cm$^{-3}$), and p-type Zn-doped ($2.94\times10^{19}$ cm$^{-3}$) GaAs. The experimental data (dots) is compared to the theoretical model (lines). The error bars are of the order of the marker size.}
    \label{fig:amplitudes}
\end{figure}

\begin{figure*}
    \centering
    \includegraphics[width=1\textwidth]{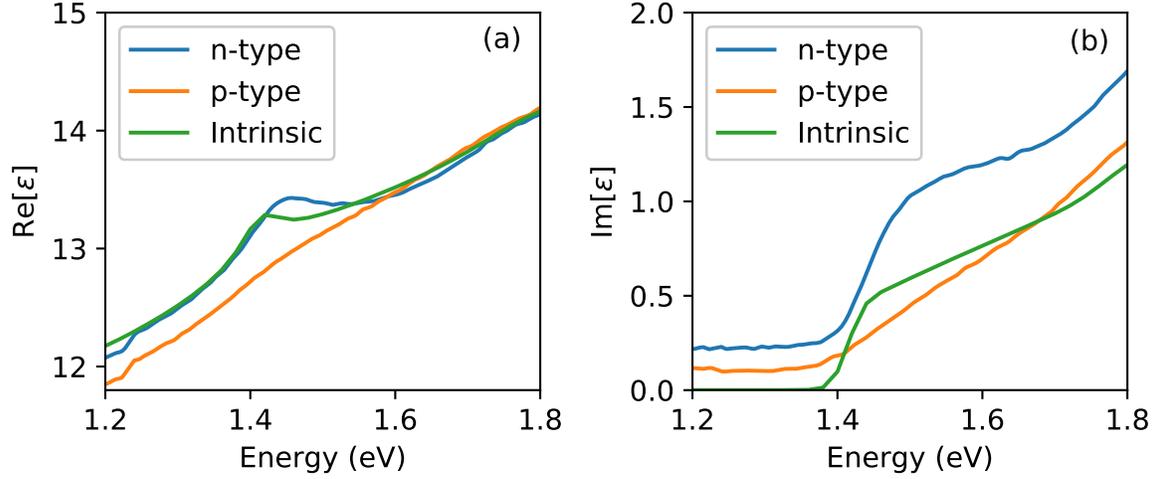}
    \caption{Real (a) and imaginary (b) parts of the dielectric function for intrinsic , n-type Te-doped ($1.28-2.35\times10^{18}$ cm$^{-3}$), and p-type Zn-doped ($2.94\times10^{19}$ cm$^{-3}$) GaAs obtained using ellipsometry.}
    \label{fig:diel_func}
\end{figure*}

The amplitude of Brillouin oscillations as a function of probe wavelength for different doping levels is plotted in Figure \ref{fig:amplitudes}. As it has been shown before \cite{Miller2006Near-bandgap-wa}, the amplitude of Brillouin oscillations changes drastically near the band edge and is maximized at the band gap ($\Gamma$ point). Such dependence can be explained by the sharpness of the band edge. It has been shown in case of GaP that the energy dependence of the amplitude of the Brillouin oscillations, $A_{osc}$, agrees well with the derivative of the dielectric function \cite{Baydin-energy-dep-2019}.

\begin{equation}
    A_{osc}\propto \left| \frac{\partial\epsilon}{\partial E}\right|=\sqrt{\left(\frac{\partial\epsilon_r}{\partial E}\right)^2
   +\left(\frac{\partial\epsilon_i}{\partial E}\right)^2},
\end{equation}
where $\epsilon$ is the dielectric function and $\epsilon_r$ and $\epsilon_i$ are the real and imaginary parts of the dielectric function, respectively. $E$ is the probe energy. Therefore, as it can be seen in Figure \ref{fig:diel_func} showing the dielectric function of GaAs, when one takes the derivative of the dielectric function, it is maximized near the band gap energy as the slope in this region is steep. To match the scale of the experimental data, all theoretical curves (the derivatives) are multiplied by the same factor which includes the photoinduced strain magnitude and other experimental parameters which are, to a good degree of approximation, constant with respect to the probe energy.

For n-type GaAs sample, the peak in the amplitude of Brillouin oscillations shifts to higher energies (shorter wavelengths) and broadens. While, for p-type GaAs sample, no dependence of the amplitude of Brillouin oscillations on probe energy is observed in the probe energy region. Note, that in our case, p-type sample has higher concentration of carriers than the n-type sample. The response of the n-type and p-type samples also agrees well with predicted energy dependence based on the derivative of the dielectric function. As dopants are added to the GaAs lattice, donor or acceptor states depending on the type of doping form near the conduction or valence bands, respectively. This formation of dopant states results in the changes in the dielectric function such as smearing of the band edge for the imaginary part of the dielectric function (see Figure \ref{fig:diel_func}b) and shifting and broadening of the peak associated with the band gap for the real part of the dielectric function (see Figure \ref{fig:diel_func}a). 
Our results demonstrate that TDBS can be used to distinguish between different dopant  concentrations. While, there are other techniques to measure doping levels, TDSB can uniquely provide depth resolution simultaneously with doping concentration. 

\begin{figure}
    \centering
    \includegraphics[width = 0.5\textwidth]{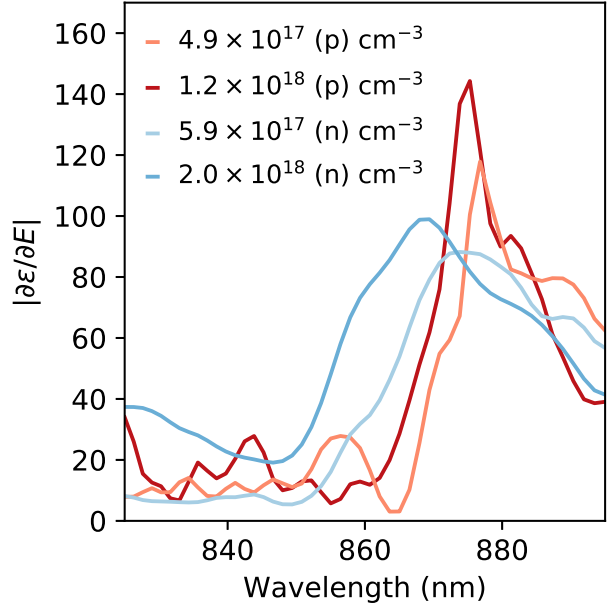}
    \caption{Derivative of the dielectric function for different doping types and concentrations. The dielectric function was taken from Casey et. al. \cite{Casey1975, sell1974concentration}.}
    \label{fig:casey}
\end{figure}

As mentioned earlier, our n-type and p-type samples have different doping concentrations. In order to get some insight how the type of the doping would affect the TBDS response, we took the dielectric functions for several doping concentrations from previous studies \cite{Casey1975, sell1974concentration}. Figure \ref{fig:casey} shows the derivative of the dielectric functions for n- and p-type GaAs at different doping concentrations. According to these theoretical estimations, the peaks for the p-type are higher and narrower than those for the n-type GaAs. This indicates that TDBS can be sensitive both to the doping concentration and the type of doping. 

\section{\label{sec:conc}Conclusion}

We have investigated the influence of type and level of doping in GaAs on Brillouin oscillations using TDBS. The experiments were carried out for intrinsic, n-type and p-type GaAs samples. The amplitude of Brillouin oscillations changes with respect to dopant level while the change in their frequency is negligible. The energy dependence of the amplitude of Brillouin oscillations is well explained by the theoretical model based on the derivative of the dielectric function. Our results show that TDBS can be used to measure dopant concentrations. This new report on the energy dependence of Brillouin oscillations adds another approach in an application of TBDS for nanoscale imaging. Particularly, we envision that TDBS spectra taken for a range of probe energies near direct optical transitions could be used to measure specific dopant concentrations as a function of depth. For example, applications  can  include  monitoring  wafer  doping  homogeneity  with  respect  to  depth,  or determining the position and interface between different doping levels and types.  As various types of defects and impurities have different effects on the dielectric function, it should be possible to distinguish them and monitor their depth distribution.  This is especially relevant for understanding the degradation of devices operating in harsh environments that are subject to radiation damage as well as for defect inspection in semiconductor wafer metrology.

\begin{acknowledgments}
The authors acknowledge the ARO for financial support under Award No. W911NF14-1-0290. Portions of this work were completed using the shared resources at the Vanderbilt Institute of Nanoscale Science and Engineering (VINSE) core laboratories.
\end{acknowledgments}

$^\dagger$ A.D. and A.B. contributed equally to this work. 

\bibliography{references_list.bib}

\end{document}